# Entrapment of Interfacial Nanobubbles on Nano-Structured Surfaces

Yuliang Wang[*,1], Xiaolai Li[1], Shuai Ren[1], Hadush Tedros Alem[1], Lijun Yang[2], and Detlef Lohse[*,3,4]

[1]School of Mechanical Engineering and Automation, Beihang University, Beijing 100191, P.R. China

[2]School of Astronautics, Beijing University of Aeronautics and Astronautics, Beijing 100191, China

[3]Physics of Fluids Group, Faculty of Science and Technology, J. M. Burgers Centre for Fluid Dynamics, University of Twente, P.O. Box 217, 7500 AE Enschede, The Netherlands

[4]Max Planck Institute for Dynamics and Self-Organization, Am Fassberg 17, 37077 Göttingen, Germany

**Abstract:** Spherical-cap-shaped interfacial nanobubbles (NBs) forming on hydrophobic surfaces in aqueous solutions have extensively been studied both from a fundamental point of view and due to their relevance for various practical applications. In this study, the nucleation mechanism of spontaneously generated NBs at solid-liquid interfaces of immersed nanostructured hydrophobic surfaces is studied. Depending on the size and density of the surface nanostructures, NBs with different size and density were reproducibly and deterministically obtained. A two-step process can explain the NB nucleation, based on the crevice model, i.e., entrapped air pockets in surface cavities which grow by diffusion. The results show direct evidence for the spontaneous formation of NBs on a surface at its immersion. Next, the influence of size and shape of the nanostructures on the nucleated NBs are revealed. In particular, on non-circular nanopits we obtain NBs with a non-circular footprint, demonstrating the strong pinning forces at the three-phase contact line.

**Keywords**: nanobubbles; nanostructured surfaces; crevice model; gas entrapment; atomic force microscope.

---

Corresponding authors:
Yuliang Wang: wangyuliang@buaa.edu.cn & Detlef Lohse: d.lohse@utwente.nl

## Introduction

When hydrophobic surfaces are immersed into aqueous solutions, spherical-cap-shaped interfacial bubbles can be obtained on these surfaces[1]. These gas nanobubbles, having heights between 5 and 100 nm and diameters between 50 nm and 800 nm, are normally referred to as surface or interfacial nanobubbles (NBs). They have great potential in numerous applications, such as drug delivery and ultrasonic tumor imaging enhancement [2], mineral flotation and separation [3, 4], or nanostructured surface fabrication [5-7]. Though it remains a challenge to distinguish them from nanodroplets and blisters [1, 8], it meanwhile has been proven that they can last for hours and even days [1, 9-11]. This stability is explained by pinning of the contact line which allows for a stable balance between gas oversaturation in the liquid and the Laplace pressure in the NB [1, 11-14]. The stable equilibrium is achieved by gas diffusion through the interface.

By definition, the formation of interfacial NBs is heterogenous nucleation. Its mechanism has been conjectured by Harvey et al. [15] and quantified by Atchley et al [16]. Basically, it is the entrapment of gas in crevices on surfaces. These bubble nuclei then can grow by diffusion or in case of strongly reduced pressure by gas expansion. Experimentally, for macro-sized bubbles, several experimental techniques have been applied to quantitatively validate this bubble nucleation model, such as a Berthelot tube, centrifugation, shock wave, acoustics, etc [17-20]. These techniques mostly focus on the reduction of liquid pressure to a negative value to find the cavitation pressure threshold. However, less studies have been performed on the nature of the crevice which is necessary to lead to NB nucleation, though on the microscale some studies were recently performed [21-23].

On a nanoscale, Checco et al. investigated the entrapment of air in 20 nm wide nanoholes on nanopatterned surfaces, employing the transmission small-angle X-ray scattering (SAXS) method [24]. By comparing the scattered X-ray peak intensities from the sample surfaces in contact with air and water, they conjectured that the water only partially enters the nanoholes for about 5-10 nm, with the rest of volume occupied by air, independent of the nanohole depth. However, the authors could not visualize the air entrapment with the SAXS. Therefore, it is not clear how exactly the water-air interface looks like. Moreover, it still remains unclear under what exact conditions air entrapment can happen.

Generally speaking, there are two approaches to make interfacial NBs[1], namely the solvent exchange method [25] and the spontaneous formation of NBs at immersion of dry samples [26]. Until now, it is still challenging to form NB in a reproducible and controllable way [1]. Among the different factors that influence NB formation[9, 27-29], surface structures must play an important role, due to their role in pinning. With nanopatterned hydrophobic/hydrophilic surfaces, Agrawal et al. [30] and Bao et al. [31] found that NBs only nucleate on hydrophobic domains. Recently, a similar study was conducted by Wang et al. on surfaces covered with nanopores [32]. Also they found that NBs will nucleate on hydrophobic areas. Although numerous efforts have been put on the topic [1], until now there is no method which can achieve fully reproducible and controllable NB formation.

In this study, we are trying to get closer to this goal of reproducible and controllable NB formation on solid-liquid interfaces by studying how NBs form on immersed nanostructured surfaces. The NB nucleation on hydrophobic surfaces with different nanostructures will be investigated. Experimental evidence of spontaneous entrapment of gas in surface nanostructures will be provided. Moreover, the study will reveal the crucial role of pinning and implies that the size, density, and even the morphology of NBs can be controlled through surface nanostructures.

## Results and Discussion

### *Nanobubbles Entrapped by Nanopits on Polystyrene Films*

The spontaneous nucleation of NBs was first studied on nanopits fabricated on spin coated polystyrene (PS) films on silicon substrates. During spin coating, the decreasing thickness of PS films will cause the inhomogeneous coating of the films [33, 34]. By delicately controlling the PS solution concentration and the spin speed during the spin coating process, nanopits of different sizes could spontaneously be obtained on two PS films (sample 1 and sample 2. For details of sample preparation, please refer to the Supporting Material). On sample 1, the width of the nanopits was in between 50-100 nm. On sample 2, their width varied from about 50 nm to 250 nm. The effect of the nanopit size on the NB entrapment was then investigated on the two samples.

The tapping mode AFM (TMAFM) image of sample 1 in air is shown in **Figure 1a**. On the surface, nanopits with typical diameter of 80 nm and depth of 4 nm are visible. The probability density function (PDF) of the nanopits' depth and width is discussed in the supporting material (see **Figure S1** in the Supporting Material). After the sample was immersed into DI water, NBs were generated by spontaneous air entrapment in the nanopits (**Figure 1b**). The nanopits were covered by NBs, such as at the location selected by the blue dashed box.

After that, NB coalescence [35-37] was performed by applying a higher scan load (TMAFM setpoint: 80% of free oscillation amplitude) in the area in DI water. Smaller NBs were moved and coalesced to generate a bigger one, as shown in **Figure 1c**. Therefore the nanopits become visible again after the coalescence process.

The coalescence of NBs makes it possible to compare the sample surface in air and in DI water with and without NBs. **Figure 1d** shows the 3D mesh plots of the same selected area in air (left), in water with an entrapped NB (middle), and in water after coalescence (right). From the figures, it is obviously that there was a NB entrapped in the nanopit.

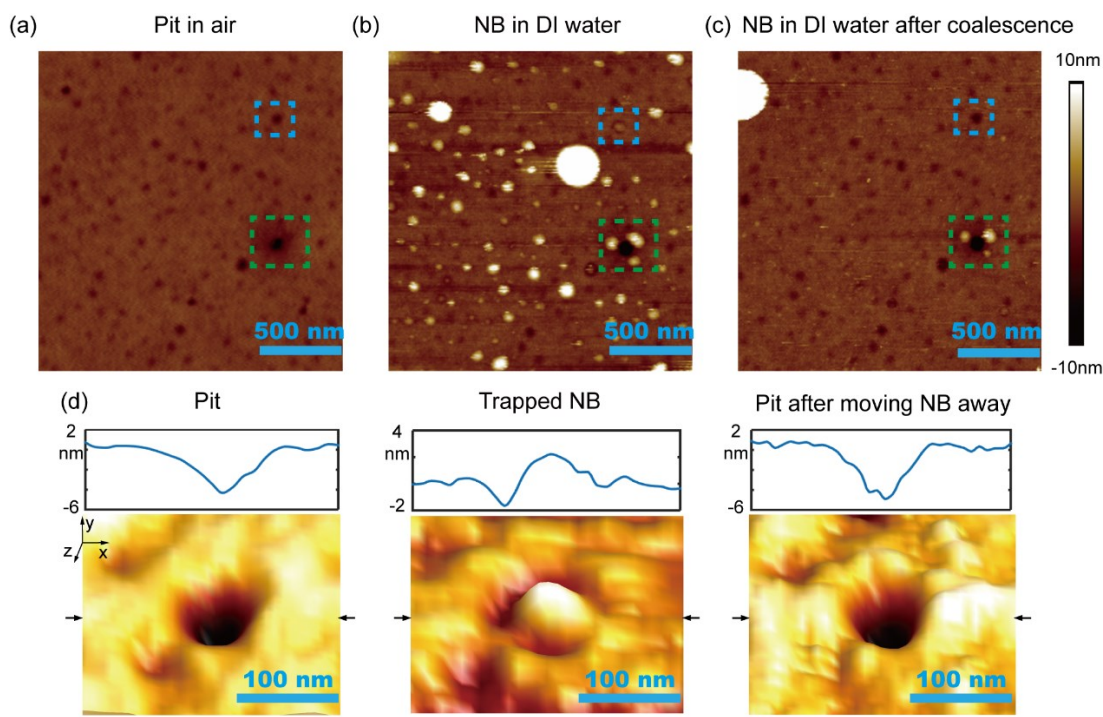

*Figure 1. Nanobubble nucleation on smaller nanopits. (a) AFM images of nanopits on a PS surface in air. On the surface, nanopits with diameter in between 50 - 100 nm are visible. (b) AFM image of the same area obtained in DI water. Nanopits can entrap NBs. (c) AFM image of the same area after NB coalescence was performed. Small NBs were coalesced to generate a bigger one (see in upper left corner of image) and the nanopits now filled with water become visible again in DI water. (d) Comparison of 3D mesh plots of the areas with a blue square selected in (a-c). One can clearly see the topography of the nanopit in air (left), an entrapped NB in the nanopit (middle), and again, the water filled nanopit (right). Note that in the area selected by a green dashed square in **Figure 1a-c**, three NBs remained after high load scan, though with reduced size compared to the ones obtained in **Figure 1b**. This is thought to be due to the existence of the large nanopit in the selected area. As reported previously, the concave surface structures help to stabilize NBs, leading to an increased NB immobility* [34]. *The enlarge 3D plot of the selected area is shown in the supporting materials (**Figure S2**).*

On sample 2, nanopits with a larger range of width were obtained. The width of nanopits varies from about 50 nm to 250 nm. The PDFs of width and depth are again shown in the Supporting Material (**Figure S3**). The influence of the nanopit size on the NB nucleation process can thus be investigated. The TMAFM image of sample 2 in air is shown in **Figure 2a**. **Figure 2b** shows the image of the sample after immersion in DI water. The surface was again covered with spontaneously generated NBs. Compared with those obtained on sample 1, the NBs obtained on sample 2 have relatively uniform size. We believe that the uniform distribution of NBs on sample 2 is related to the spin coating speed during sample preparation. Sample 2 was made with a high spin coating speed, which can lead to an uniform roughness, thus resulting in uniform nucleation of NBs.

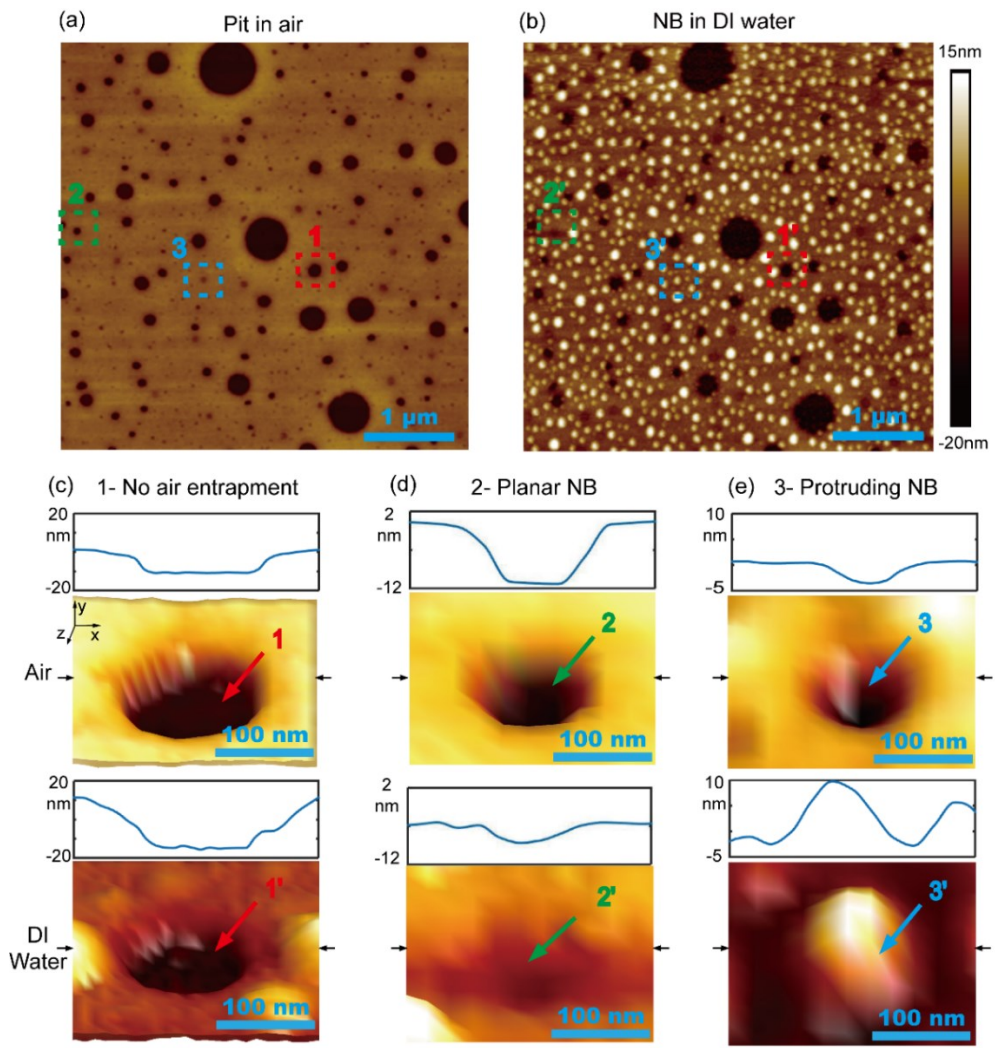

*Figure 2. Entrapped nanobubbles on larger nanopits. (a) TMAFM image of nanopits with*

*different widths on a PS film in air. (b) AFM image of the same area obtained in DI water. NBs were spontaneously generated in some nanopits on the PS film after the sample was immersed in DI water. (c-e) Comparison of three selected nanopits of different sizes in air (top row) and in DI water (bottom row) with cross section profiles. (c) A wide nanopit with no NB entrapped. (d) A medium nanopit with an entrapped planar NB. (e) A narrow nanopit with an entrapped NB of protruding profile.*

One can see that there are now three types of air entrapment on sample 2, as marked by 1, 2, and 3 in **Figure 2a** and **2b**. The mesh plots of the three selected areas are given in **Figure 2c, 2d, 2e**, respectively. The first one is for the nanopits with width larger than 150 nm, as shown in **Figure 2c**. The width and depth of the nanopit in water are the same as that in air, which means there is no air entrapped in the pit. The second one is for the nanopits with width in between 100 nm and 150 nm, as shown in **Figure 2d**. On this kind of nanopits, *planar* NBs can be observed. From the cross section of the top AFM image, the measured depth of the nanopit in air was about 15 nm. However, the cross section of the profile becomes relatively flat in DI water, which is believed to be the entrapped air. The third case is for nanopits of less than 100 nm in width, as demonstrated in **Figure 2e**. On these nanopits, regular NBs with higher contact angle are generated, similarly as already done on sample 1 above, with similar nanopit extensions. In this study, these NBs are referred to as protruding NBs.

The classification of planar and protruding NBs is based on the value of the contact angle. The contact angles of the nanobubbles were determined from the NB cross sections. Assuming that the NBs are spherical-cap-liked objects, their contact angles can be calculated by measuring their height and footprint diameter. However, it is well known that a topography image obtained from an AFM is actually the convolution of the AFM tip and substrate morphologies [38-40]. Therefore, the measured NB width needs to be corrected due to the finite size of the cantilever tip radius. In this study, the details of AFM tip deconvolution are included in the supporting information (**Figure S7a**). For the NBs with width of 50-150 nm, the result shows that $D$ is about 2-4 nm larger than $D'$. This results in about 2% underestimation of NB contact angle, as shown in a comparison of the corrected and uncorrected contact angles for protruding NBs in the supporting information (**Figure S7b**).

With the corrected NB contact angles, the trapped NBs are classified into two groups (please refer to **Figure S8** for the detailed distribution of contact angles for trapped NBs). In this study, the NBs with contact angles of 0º or even slightly less are referred to as planar NBs. The others are protruding NBs, of which contact angles are normally larger than 10 º.

To get a better understanding of the profile for the *planar* NBs, the cross sections for all the *planar* NBs in **Figure 2** are presented in **Figure 3a**. The profiles are mostly slightly concave. The blue thick curve in **Figure 3a** is the average one of all the cross sections of the same size, from which it is clear that the cross section is slightly concave. The height $H$ of planar NBs is thus slightly negative, and correspondingly also the contact angle. The measured contact angles for all planar NBs are shown in **Figure 3b**, with an average value of -3.7º±2.5º.

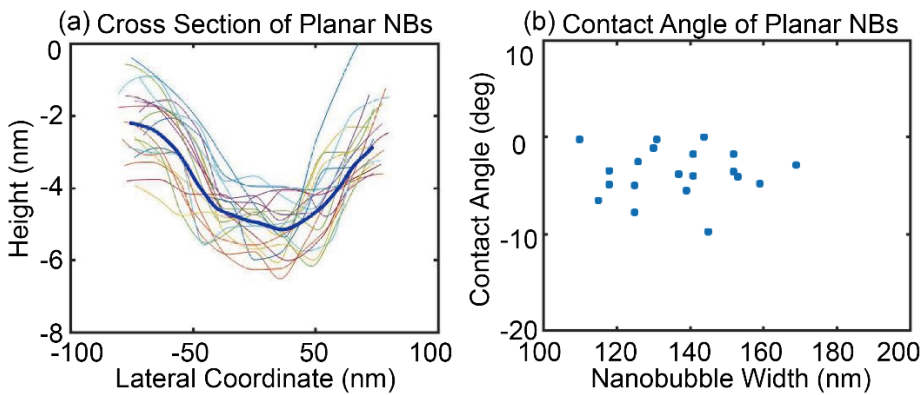

*Figure 3. Geometric analysis of planar NBs.* *(a) Cross sections of all planar NBs in Figure 2. Obviously, the real bubble shape corresponds to a minimal surface (spherical-cap) and the wiggles originate from AFM measurement noise. The blue thick curve is the average one of all the cross sections of NBs with the same lateral range from*

*-70 nm to 70 nm in the figure, which indicates slightly concave cross sections of the nearly planar NBs. (b) Contact angle of all the planar NBs with an average value of -3.7º±2.5º.*

From the above result, one can see that the types of NB nucleation are directly related to nanopit sizes. Air entrapment preferably occurs on nanopits with smaller diameter. Moreover, the nucleation on nanopits can be divided into three categories: no entrapment, *planar* NBs, and protruding NBs.

For all NBs, we counted the numbers of NBs entrapped in nanopits on sample 1 and 2. A histogram of the three nucleation categories as a function of nanopit width is shown in **Figure 4a**. The histograms of these three kinds of NB entrapment are fitted with normal distributions (dashed lines) in **Figure 4a**.

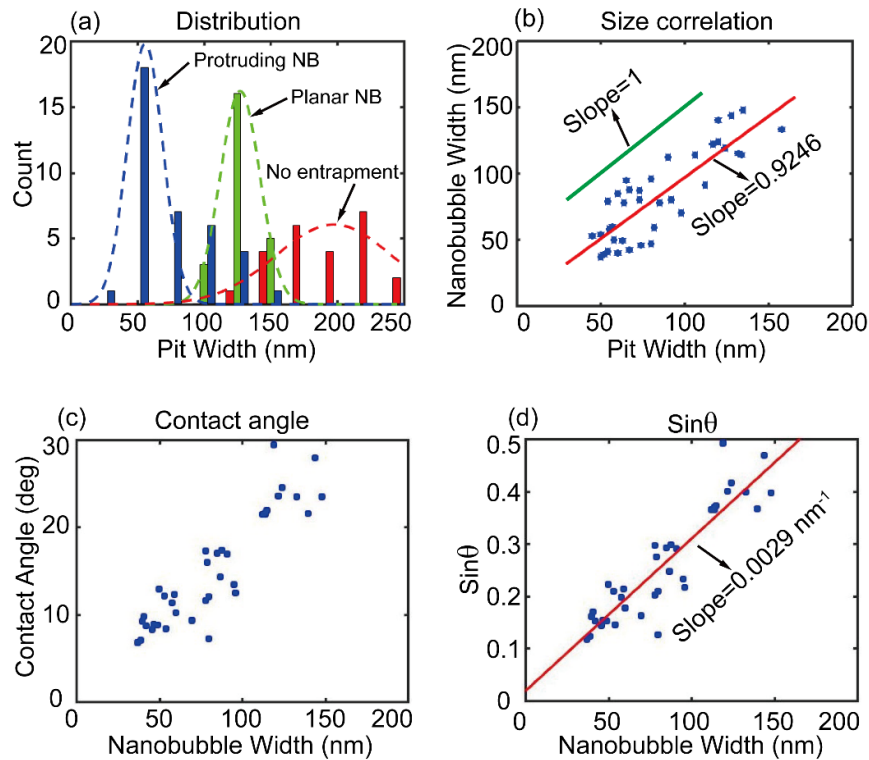

*Figure 4 Statistic analysis of nanobubble entrapment on nanopits. (a) Histogram of the three different categories of NB nucleation in DI water. The protruding NBs and planar NBs preferably form on nanopits of 50-100 nm and 100 -150 nm, respectively. However,*

*when nanopits are larger than 150 nm, there are no bubbles entrapped. (b) Size correlation between the entrapped protruding NBs and nanopits. The size of nanobubbles linearly increases with that of nanopits. (c) Contact angle of NBs as a function of NB width for the protruding NBs. The contact angle increases with increasing NB width. (d) Sinθ vs nanobubble width L. According to Eq (4), the slope is ζ/L$_c$ = 2.9 μm$^{-1}$, corresponding to a gas oversaturation of ζ = 8.2.*

For all nanopits on the two samples, the results clearly show that the protruding NBs preferably form on smaller nanopits with medium size of 60 nm. For nanopits with width in between 100 nm and 150 nm, *planar* NBs are likely to be nucleated. For nanopits with width larger than 150 nm, generally no NBs were entrapped.

Moreover, experimental results show that the size of the protruding NBs is directly related to that of nanopits. **Figure 4b** shows the correlation between the width of NBs and that of the nanopits for the protruding NBs. The data were fitted with a linear relation with the least square method. The slope of the fitted line is 0.9246, which is close to 1. This indicates that the size of the entrapped NBs is basically the same as that of nanopits.

Additionally, the contact angle (gas side) of the entrapped NBs is in between 5° and 30° and increases with the NB width, as shown in **Figure 4c**. We hypothesize that these small NBs are in diffusive equilibrium. For NBs in diffusive equilibrium, the stable equilibrium contact angle $θ_e$ is determined by the air oversaturation ζ through the equilibrium condition [1, 13],

$$\sin θ_e = ζ \frac{L}{L_c} \qquad (1)$$

Here $L$ is the pinning length and $L_c = 4σ/P_0 = 2.84$ μm, following from surface tension σ and ambient pressure $P_0$. The relation between $\sinθ_e$ and the NB width $L$ is given in **Figure 4d**. Again, the data were fitted with a linear relation with the least square method. From the linear fit, we obtain ζ/$L_c$=2.9 μm$^{-1}$, corresponding to a (local) air oversaturation of ζ =8.2, which does not seem unrealistic to us.

The process of NB formation on nanopits can be thought of consisting of two steps, as illustrated in **Figure 5**. The first step is air entrapment in surface cavities. Since PS is hydrophobic, the water will have a large advancing contact angle (**Figure 5a**) on the PS films. During the immersion, as the fluid advances down to the bottom of the nanopits, by considering the narrow lateral size of the nanopits and high advancing contact angle, an air pockets will be entrapped in the nanopits (**Figure 5b** and **Figure 5c**). It is due to air entrapment in hydrophobic conical [16, 41] and some other shapes of cavities [42], or in micro-protrusions of immersed superhydrophobic surfaces [43-45], i.e., the crevices. During entrapment, the water has to retain its position at the pit mouth until the advancing contact angle of the fluid flow is obtained. As a result, nanopits with narrow lateral length will get air entrapped inside and form an initial gas-liquid interfaces. At this stage, the curvature of the advanced fluid will be towards the entrapped air direction. The process obviously depends on the degree of hydrophobicity, i.e., on the advancing contact angle. This is consistent with the optical observation of the motion of the three-phase contact line on micropillar decorated superhydrophobic surfaces [46].

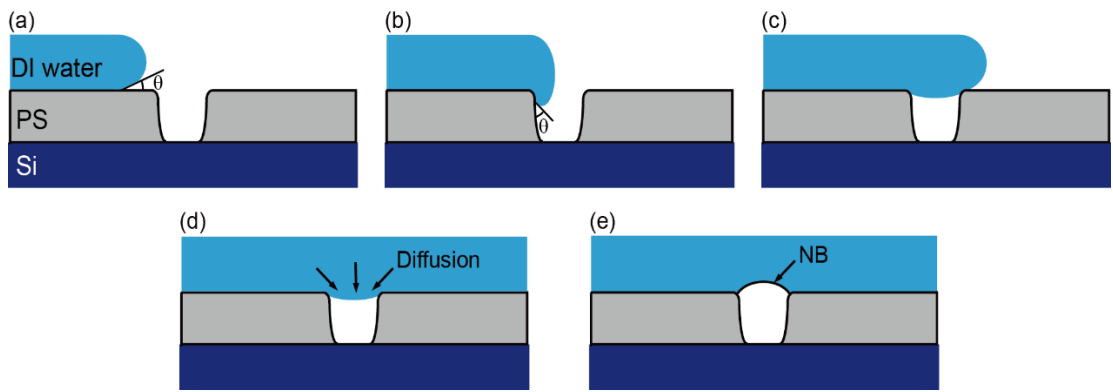

*Figure 5. Illustration of a two-step formation of NBs. (a-c) Formation of air pockets in surface cavities. (d-e) Air diffusion through liquid-air interfaces.*

The second step is the diffusive volume adjustment of NB originating from the entrapped air inside the nanopit. It is well known that the diffusion of gas from bubbles to fluid and vice versa plays an important role in NB dynamics and NB stability [1, 11, 13]. In
11

this second step, the NB will grow out from the nanopits due to the diffusion of the air out of the locally oversaturated water (**Figure 5d**), provided of course it is oversaturated. The curvature of the liquid interface inside the nanopit will start changing its direction outwards. Thanks to the diffusive influx, the liquid-air interface will grow out of the cavities until it reaches its equilibrium value determined by the air oversaturation in water. For nanopits with larger width, it might need much higher diffusive influx and the meniscus cannot so quickly appear above the nanopit mouth, i.e., diffusive equilibrium may not be achieved.

This two-step NB nucleation model is consistent with the experimental observation. Based on the experimental results, a schematic is presented to illustrate the air entrapment on nanostructured surfaces (**Figure 6**). There are three different cases, as shown in **Figure 6a**, **6b**, and **6c**, respectively. For nanopits with a width of about 50 nm, NBs with protruding profiles can be obtained (**Figure 6a**). For these NBs, the size linearly increases with the nanopit width and in diffusive equilibrium the contact angle increases with increasing NB width, according to Eq. (1).

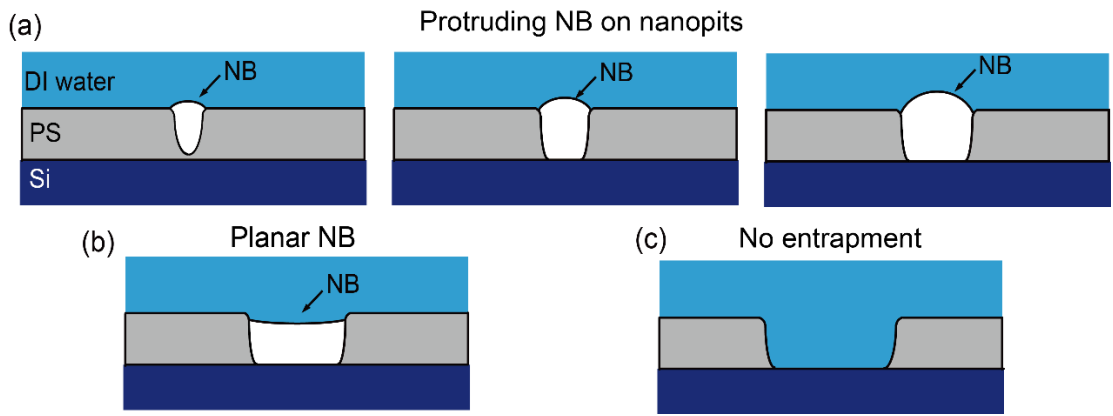

*Figure 6. Illustration of three categories of air entrapment on nanopits. (a) Protruding NBs on smaller nanopits. The size of the entrapped NBs linearly depends on that of the nanopits and NB contact angle increases with increasing NB size. (b) Planar NBs with a concave gas-liquid interfaces formed in medium sized nanopits. (c) No entrapment. For larger nanopits, gas cannot be entrapped and there are no NBs observed.*

When the nanopit width is in the range of 100-150 nm, entrapped air with flat liquid-gas interface can be obtained (**Figure 6b**). Although these nanopits also entrapped air, the liquid-gas interface does not grow out of the nanopits. Regarding the reason to the formation of *planar* NBs, one reason could be that they are not in diffusive equilibrium yet. The diffusive timescale [1] is $\tau_D \sim L^2\rho/Dc_s\zeta$, i.e., $\propto L^2$, i.e., for larger pits it takes much longer to equilibrate. In addition, for these medium sized nanopits, it requires more oversaturated air to make the gas-liquid interfaces grow out of the nanopits and depletion effects may occur. For even larger nanopit size beyond 150 nm, based on the crevice model, the nanopits do not favor air entrapment and no NB nucleate (**Figure 6c**).

*Are nanobubbles always spherical-cap-shaped?*

The experimental results obtained from sample 1 and sample 2 reveals that the size and position of the formed NBs could be tuned by modifying the nanostructures. NBs are normally believed to be of spherical-cap-shaped, as that shape is energetically the most effective one. However, stable NBs also require pinning and thus their shape may also be tuned by that of the surface nanostructures. Indeed, by surface energy minimization, Dević et al. showed how chemical heterogeneities determine the shape of nanodroplets [47]. For a non-circular chemical or geometric surface pattern, it may indeed be energetically advantageous for a NB or nanodrop (ND) to have a non-circular footprint.

In this study, an ultrathin PS film (sample 3) was used to generate a porous surface with irregular shapes of pores. The nanopores were fabricated with assistance of NBs. As previous revealed [5], when an ultrathin PS film is immersed in DI water, small NBs first disappear and leave nanopores on the surface. The formation of nanopores is due to surface forces along the three-phase contact line of the NBs and high inner pressure. With the decreased PS film thickness (ultrathin one, in this case), its elastic modulus and fracture strength will be reduced as compared to the bulk values. The interaction between the film and the NBs will disentangle the PS chain and lead to the formation of irregular shapes of nanopores.

A TMAFM image of the porous surface fabricated with assistance of NBs is shown in **Figure 7a**. Unlike the circular shapes of the nanopits shown on sample 1 and sample 2, here the nanopores on sample 3 have irregular shapes. For the detailed process of the formation of nanopores, please refer to the supporting information.

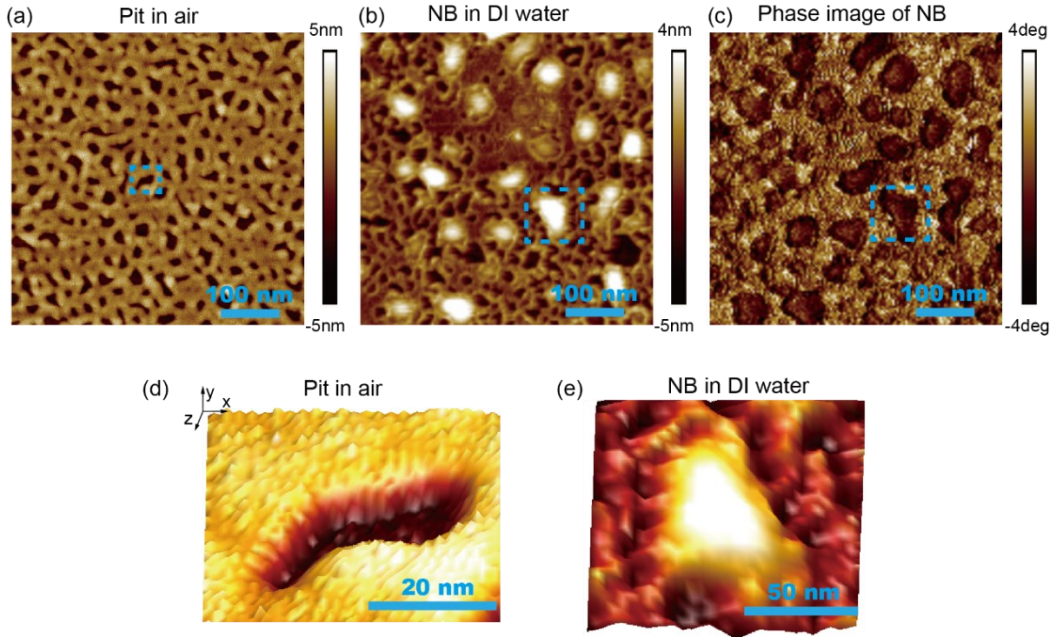

*Figure 7 Non-spherical-cap-shaped nanobubbles entrapped on a porous PS surface covered with nanopores of irregular shapes. (a) TMAFM height image of the PS sample surface covered with nanopores of irregular shapes. (b) TMAFM height image of NBs on the surface after immersed into DI water. (c) TMAFM phase image of the same scan area. The dark areas indicate the footprint of NBs. It is clear that most NB footprints are not circular, i.e., the NBs have a non-spherical-cap-shape. (d) A 3D mesh plot of a selected area in (a), which shows the irregular shape of the nanopore. (e) A 3D mesh plot of a selected area in NB height image of (b), which shows a non-spherical-cap-shaped NB nucleated on the sample.*

After sample 3 was immersed into DI water, NBs were formed on the surface. The height image is shown in **Figure 7b**. From the figure, we can see that the shape of the NBs is very different as compared to the regular NBs on the circular pits. This can be better seen in the phase image **(Figure 7c)**, which clearly shows that the shape of NBs is no longer circular. **Figure 7d** and **7e** show 3D mesh plot of selected areas in **Figure 7a** and **7b**, respectively. In the 3D images, the nanohole is no longer circular and the NB is not spherical-cap-shaped any more, but shows an irregular shape.

The existence of stable non-spherical-cap-shaped NBs highlights the relevance of pinning for NB formation, namely here at the edge of the irregular surface cavities on sample 3. On regular smooth sample surfaces with rms (root mean square) roughness about 0.2 nm [37] or on the circular nanopits shown on sample 1 and 2, NBs are spherical-cap-shaped. Indeed, the shape of the NBs is determined by two factors. One is the interfacial free energy of the NB, the other is the constraint from the pinning of the three-phase contact line of the NB. The pinning forces can be considerable. Tan *et al.*[48] found that a pinning force up to 0.1 $\mu$N is required to unpin bubbles of several micrometers, which corresponding to 5 mN/m unpinning force to detach bubbles. Similarly, as in above mentioned numerical study of NBs and NDs on chemically heterogenous surfaces [47], also here the surface geometrical heterogeneity plus the minimized interfacial free energy determine the shape of the NBs. On sample 3, the three-phase contact line of NBs can only lie on the ridge area of PS (see **Figure 7a**). In other words, the pinning sites provided by the ridge area, together with the minimized free interfacial energy, determine the shape of the NBs.

The non-spherical-cap-shaped NBs indicate that the NB footprint is directly related to the corresponding surface nanostructure to which it is pinned. NBs can change their shape to adapt to that of the surface nanostructures on substrates and they do not necessarily be spherical-cap-shaped. This implies that not only the size and locations of the NBs, but also the shape of NBs can be tuned by nanostructures.

## Conclusion

In summary, the spontaneous nucleation of NBs on nanostructured surfaces was investigated. Three categories of air entrapment were observed, namely protruding NBs on nanopits with widths in between 50-100 nm, *planar* NBs on nanopits with widths of 100-150 nm, and no air entrapment on nanopits with diameter larger than 150 nm. The size of the entrapped NBs linearly increases with that of nanopits. A two-step nucleation process was used to explain the entrapment of NBs on surface nanocavities. Air is first entrapped by surface cavities during immersion with the formed liquid-air interfaces. After that, air dissolved in the locally over-saturated liquid diffuses into the air pockets. Then the air pockets grow out of the cavities, determining the final volume of the NBs. Moreover, air entrapment experiments on a porous PS surface covered with irregular nanopores show that also the shape of NBs can also be tuned by surface nanostructures, due to their pinning effect. It is thus feasible to tune NB size, position, and even morphology by varying surface nanostructures.

**Acknowledgement** D. L. thanks Xuehua Zhang for numerous illuminating discussions over the years. This study is financially supported by National Natural Science Foundation of China (Grant No. 51305019) and Beijing Natural Science Foundation (Grant No. L150002).

**References:**

1. D. Lohse and X. Zhang, *Rev. Mod. Phys.*, 2015, **87**, 981-1035.
2. Z. Xing, J. Wang, H. Ke, B. Zhao, X. Yue, Z. Dai and J. Liu, *Nanotechnology*, 2010, **21**, 145607.
3. A. Azevedo, R. Etchepare, S. Calgaroto and J. Rubio, *Miner. Eng.*, 2016, **94**, 29-37.
4. S. Liu, J. Duvigneau and G. J. Vancso, *Eur. Polym. J.*, 2015, **65**, 33-45.
5. Y. Wang, B. Bhushan and X. Zhao, *Nanotechnology*, 2009, **20**, 045301.
6. C. Huang, J. Jiang, M. Lu, L. Sun, E. I. Meletis and Y. Hao, *Nano Lett.*, 2009, **9**, 4297-4301.
7. L. Gao, G.-X. Ni, Y. Liu, B. Liu, A. H. Castro Neto and K. P. Loh, *Nature*, 2014, **505**, 190-194.
8. R. P. Berkelaar, E. Dietrich, G. A. M. Kip, E. S. Kooij, H. J. W. Zandvliet and D. Lohse, *Soft Matter*, 2014, **10**, 4947-4955.
9. J. W. Yang, J. M. Duan, D. Fornasiero and J. Ralston, *J. Phys. Chem. B*, 2003, **107**, 6139-6147.
10. X. H. Zhang, A. Quinn and W. A. Ducker, *Langmuir*, 2008, **24**, 4756-4764.
11. X. Zhang, D. Y. C. Chan, D. Wang and N. Maeda, *Langmuir*, 2013, **29**, 1017-1023.
12. Y. Liu and X. Zhang, *J. Chem. Phys.*, 2014, **141**, 134702.
13. D. Lohse and X. Zhang, *Physical Review E*, 2015, **91**, 031003.
14. J. H. Weijs and D. Lohse, *Phys. Rev. Lett.*, 2013, **110**, 054501.
15. E. N. Harvey, D. K. Barnes, W. D. Mcelroy, A. H. Whiteley, D. C. Pease and K. W. Cooper, *Journal of Cellular & Comparative Physiology*, 1944, **24**, 1-22.

16. A. A. Atchley and A. Prosperetti, *J. Acoust. Soc. Am.*, 1989, **86**, 1065-1084.
17. B. M. Borkent, S. Gekle, A. Prosperetti and D. Lohse, *Phys. Fluids*, 2009, **21**, 102003.
18. N. Bremond, M. Arora, C. D. Ohl and D. Lohse, *Phys. Rev. Lett.*, 2006, **96**, 224501.
19. F. Caupin and E. Herbert, *C. R. Phys.*, 2006, **7**, 1000-1017.
20. E. Herbert, S. Balibar and F. Caupin, *Physical Review E*, 2006, **74**, 041603.
21. X. Zhang, H. Lhuissier, O. R. Enriquez, C. Sun and D. Lohse, *Langmuir*, 2013, **29**, 9979-9984.
22. O. R. Enriquez, C. Sun, D. Lohse, A. Prosperetti and D. van der Meer, *J. Fluid Mech.*, 2014, **741**, R1-9.
23. S. R. German, X. Wu, H. An, V. S. J. Craig, T. L. Mega and X. Zhang, *ACS Nano*, 2014, **8**, 6193-6201.
24. A. Checco, T. Hofmann, E. DiMasi, C. T. Black and B. M. Ocko, *Nano Lett.*, 2010, **10**, 1354-1358.
25. S. T. Lou, Z. Q. Ouyang, Y. Zhang, X. J. Li, J. Hu, M. Q. Li and F. J. Yang, *J. Vac. Sci. Technol. B*, 2000, **18**, 2573-2575.
26. N. Ishida, T. Inoue, M. Miyahara and K. Higashitani, *Langmuir*, 2000, **16**, 6377-6380.
27. R. P. Berkelaar, J. R. T. Seddon, H. J. W. Zandvliet and D. Lohse, *ChemPhysChem*, 2012, **13**, 2213-2217.
28. X. Cui, C. Shi, L. Xie, J. Liu and H. B. Zeng, *Langmuir*, 2016, **32**, 11236-11244.
29. C. Xu, S. Peng, G. G. Qiao, V. Gutowski, D. Lohse and X. Zhang, *Soft Matter*, 2014, **10**, 7857-7864.
30. A. Agrawal, J. Park, D. Y. Ryu, P. T. Hammond, T. P. Russell and G. H. McKinley, *Nano Lett.*, 2005, **5**, 1751-1756.
31. L. Bao, Z. Werbiu, D. Lohse and X. Zhang, *J. Phys. Chem. Lett.*, 2016, **7**, 1055-1059.
32. L. Wang, X. Y. Wang, L. S. Wang, J. Hu, C. L. Wang, B. Y. Zhao, X. H. Zhang, R. Z. Tai, M. D. He, L. Q. Chen and L. J. Zhang, *Nanoscale*, 2017, **9**, 1078-1086.
33. G. Reiter, *Langmuir*, 1993, **9**, 1344-1351.
34. Y. Wang, B. Bhushan and X. Zhao, *Langmuir*, 2009, **25**, 9328-9336.
35. B. Bhushan, Y. Wang and A. Maali, *J. Phys.: Condens. Matter*, 2008, **20**, 485004.
36. Y. Wang, H. Wang, S. Bi and B. Guo, *Beilstein Journal of Nanotechnology*, 2015, **6**, 952-963.
37. Y. Wang, H. Wang, S. Bi and B. Guo, *Scientific Reports*, 2016, **6**, 30021.
38. B. M. Borkent, S. de Beer, F. Mugele and D. Lohse, *Langmuir*, 2010, **26**, 260-268.
39. D. Li and X. Zhao, *Colloid Surf. A-Physicochem. Eng. Asp.*, 2014, **459**, 128-135.
40. B. Song, W. Walczyk and H. Schoenherr, *Langmuir*, 2011, **27**, 8223-8232.
41. M. A. Chappell and S. J. Payne, *Respir. Physiol. Neurobiol.*, 2006, **152**, 100-114.
42. M. A. Chappell and S. J. Payne, *J. Acoust. Soc. Am.*, 2007, **121**, 853-862.
43. T. M. Schutzius, S. Jung, T. Maitra, G. Graeber, M. Koehme and D. Poulikakos, *Nature*, 2015, **527**, 82-85.
44. F. Geyer, C. Schoenecker, H.-J. Butt and D. Vollmer, *Adv. Mater.*, 2017, **29**, 1603524.
45. P. Papadopoulos, D. Vollmer and H.-J. Butt, *Phys. Rev. Lett.*, 2016, **117**, 046102.
46. F. Schellenberger, N. Encinas, D. Vollmer and H. J. Butt, *Phys. Rev. Lett.*, 2016, **116**, 096101.
47. I. Devic, G. Soligno, M. Dijkstra, R. van Roij, X. Zhang and D. Lohse, *Langmuir*, 2017, **33**, 2744-2749.
48. B. H. Tan, H. An and C.-D. Ohl, *Phys. Rev. Lett.*, 2017, **118**, 054501.

# Entrapment of Interfacial Nanobubbles on Nano-Structured Surfaces

Yuliang Wang[*,1], Xiaolai Li[1], Shuai Ren[1], Hadush Tedros Alem[1], Lijun Yang[2], and Detlef Lohse[*,3,4]

[1]School of Mechanical Engineering and Automation, Beihang University, Beijing 100191, P.R. China
[2]Physics of Fluids Group, Faculty of Science and Technology, J. M. Burgers Centre for Fluid Dynamics, University of Twente, P.O. Box 217, 7500 AE Enschede, The Netherlands
[3]Max Planck Institute for Dynamics and Self-Organization, Am Fassberg 17, 37077 Göttingen, Germany
[4]School of Astronautics, Beijing University of Aeronautics and Astronautics, Beijing 100191, China

## Methods

### *Sample preparation*

The polystyrene (PS) surfaces used for nanobubble (NB) nucleation were prepared by spin coating thin films of PS on silicon (100) substrates. Before spin coating, the substrates were sequentially cleaned in sonication bathes of piranha, acetone, and then water. PS particles (molecular weight 350 000, Sigma-Aldrich) were dissolved in toluene (Mallinckrodt Chemical) to make the PS solutions.

Three PS surfaces (sample 1, sample 2, and sample 3) were prepared to obtain nanopits with different sizes and shapes. The PS concentrations for the three samples were 0.05%, 0.1%, and 0.05% (all weight), respectively. The selected speeds of spin coating for the three samples were 200 rpm, 1000 rpm, and 500 rpm, respectively.

Two types of nanostructures, namely nanopits and nanopores, were fabricated on the three samples to study NB nucleation. First, on *nanopits*: For spin coated PS films, the decreasing thickness will cause inhomogeneous coating of the films[1, 2]. By delicately controlling the PS solution concentration and spin speed during spin coating, nanopits of different sizes were spontaneously obtained on sample 1 and sample 2.

On sample 3, *nanopores* were fabricated with assistance of NBs, as first reported in our previous study [3] and later by Janda et al [4]. When PS films were immersed in water, small bubbles disappear with time and generate nanopores on sample surfaces [3]. Similar results were also

Corresponding authors:
Yuliang Wang: wangyuliang@buaa.edu.cn & Detlef Lohse: d.lohse@utwente.nl

reported on graphene surfaces [5]. Recently, by applying negative pressure, Janda et al. obtained nano pinholes and nano protrusions, respectively, on thinner and thicker PS films, respectively. The NB caused generation of nanostructures on PS surfaces is mostly due to the presence of surface forces at the three phase contact line and the high inner pressure of NBs [6].

Through experiments, we found that the type of NB assisted nanostructures is strongly related to two factors, namely the NB size and the thickness of the PS films. Nanostructures tend to form namely on ultrathin PS films (with thickness from 3 nm to 7 nm). This is mostly because the polymer chain entanglement density is greatly reduced with decreasing sample thickness, as motivated theoretically [7, 8] and shown experimentally [9, 10]. This will lead to a much lower elastic modulus and fracture strength of ultrathin PS films as compared to their bulk values. During our experiments, smaller NBs (less than 50 nm in diameter and thus with much higher inner pressure, assuming constant contact angle) mostly disappear and lead to formation of nanopores on ultrathin PS film with thickness around 3 nm. Bigger NBs (larger than 80 nm) last longer and will lead to the formation of nanoindents, even on a relative thick PS film.

Sample 3 was first immersed into DI water and smaller NBs (with width less than 50 nm) were formed on the surface. After that, NBs disappeared and a porous PS surface was obtained. Water was then removed from the sample and the sample was kept in an vacuum chamber for 4 h to remove the remaining water. The experiment in fluid was performed to study NB nucleation.

*AFM Measurement*

A commercial AFM (Resolve, Bruker) operating in tapping mode was used for imaging the sample in both air and deionized (DI) water. A silicon cantilever (NSC36/ALBS, MikroMasch) with a quoted tip radius of 8 nm and stiffness of 0.6 N/m from the manufacturer was used for the scanning in air and DI water. The measured resonance frequencies of the cantilever in air and water were about 55 KHz and 16 KHz, respectively. While imaging in air and liquid, the drive frequencies were slightly lower than that of the selected resonance. A scan rate of 2 Hz with a 0° scan angle was used. All experiments were performed at an ambient environment (temperature: $26 \pm 1$ °C).

In all three experiments, the sample surfaces were first scanned in air by tapping mode AFM (TMAFM). Then they were immersed into DI water to perform liquid imaging. The image scanning began immediately within 10 minutes after immersing the surface into water. For sample 1, the scanning lasted for more than 5 hours. For sample 2, the scanning lasted for more than 10 hours. The free oscillation amplitude of the cantilever at working frequency was about 290 mV. The larger setpoint oscillation amplitude of about 280 mV (namely 96.5% of free amplitude) was applied for liquid imaging to minimize disturbance of AFM tips to NBs.

As mentioned above, nanopits of different sizes were obtained on PS films. For sample 1, the distributions of the nanopits' width and depth are shown in **Figure S1(a)** and **S1(b)**, respectively. The nanopit width distributed from 50 nm to 100 nm. The nanopit depth mostly changed in between 1 to 6 nm. There was one nanopit with a depth of 12 nm, which reached down to the silicon substrate. Figure S1(a) shows the correlation between width and depth.

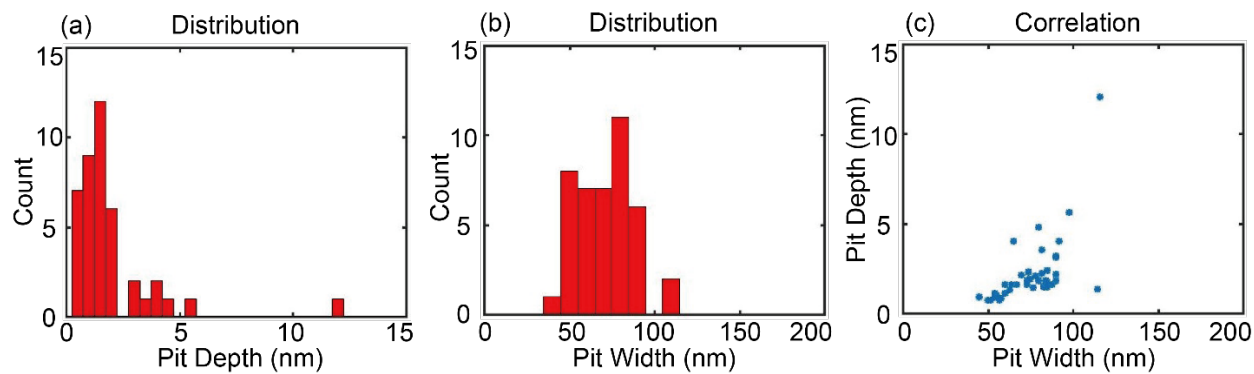

*Figure S1 Width and depth distribution of nanopits on sample 1. (a) Histogram of nanopit widths on sample 1. The nanopit width changed from 40 nm to 110 nm. (b) Histogram of nanopit depths on sample 1. The nanopit depths distributed from 1 nm to 6 nm. The correlation between width and depth is shown in (c).*

In the analysis of **Figure 1**, we found that three NBs remained on their initial locations around a large nanopits after NB coalescence. In the area selected by a green dashed box in **Figure 1(c)**, one can see that the three NBs remained after NB coalescence. **Figure S2** shows the zoom-in mesh plots of the selected area in **Figure 1(a-c)**. As mentioned in the main manuscript, this is thought to be due to the existence of the large nanopit in the selected area, providing strong pinning sites.

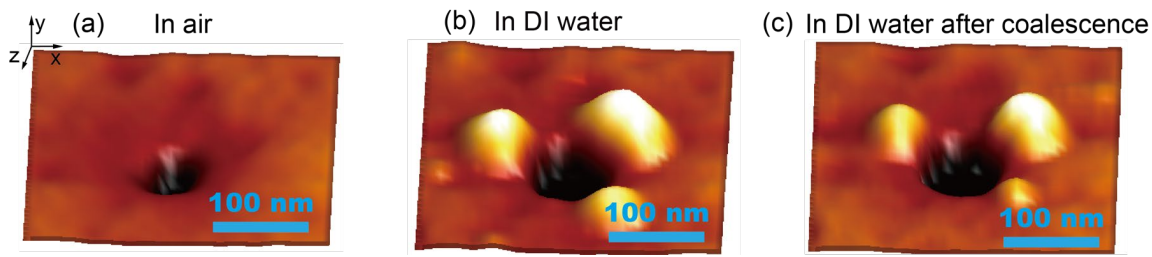

*Figure S2 A deep and large nanopit on sample 1. 3D mesh plots of the same areas selected in Figure 1. One can clearly see the topography of the nanopit in air (a), in DI water (b), and after coalescence (c).*

For sample 2, the distributions of nanopits' width and depth are shown in **Figure S3(a)** and **S3(b)**, respectively. On the sample, one can see that the nanopit widths mainly distributed from 50 nm to 250 nm. The nanopit depths distributed in two sections, 3-4 nm, and 10-13 nm. For the nanopits of which depth located in the section section (10-13 nm), they reached down to the bottom of the silicon substrate.

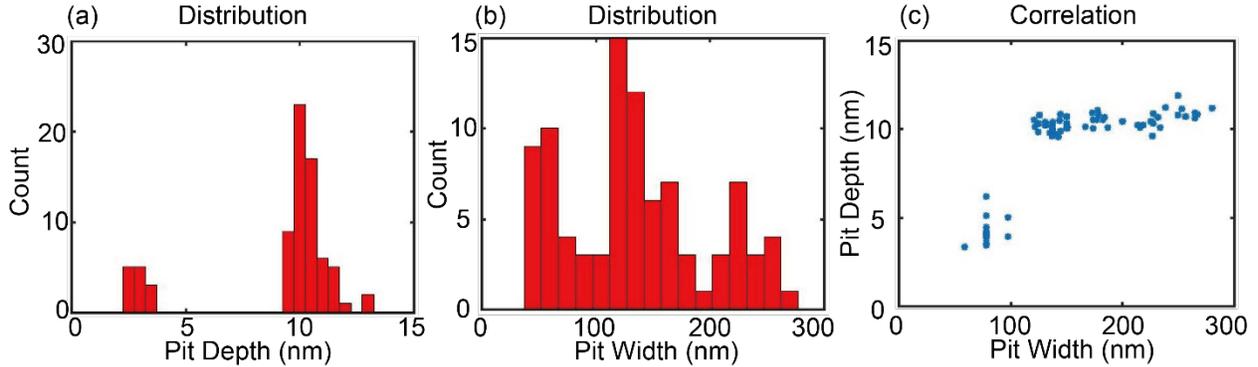

*Figure S3 Width and depth distribution of nanopits on sample 2. (a) Histogram of nanopit widths on sample 2. The nanopit widths mainly changed from 50 nm to 250 nm. (b) Histogram of nanopit depths on sample 2. The nanopit depths distributed in two sections, 3 – 4 nm, and 10-13 nm. (c) Corrrelation of width and depth.*

## Nanobubble Verification Experiment

Studies on surface NBs have been suffering from artifacts for years as explained in [11, 12]. For the samples used in this study, there are potentially three kinds of artifacts, namely solid particles, liquid objects [12], and blisters [13]. To value out such artifacts, several experiments were designed and conducted to verify that the spherical objects we observed were indeed NBs. In this supporting information, we report the result of several verification experiment.

### *Nanobubble Coalescence Experiment*

The nanobubble coalescence experiment was performed to distinguish nanobubbles from solid particles and blisters. It is well known that nanobubbles can move and coalesce to generate larger ones under higher scan load [1, 14, 15]. However, blisters and solid objects cannot coalesce. In this study, we performed NB coalescence experiment on PS samples. The result for one of these experiments is show in **Figure S4**. **Figure S4(a)** is a height image of the sample surface obtained with 96% setpoint value in TMAFM after the sample was immersed into DI water. One can see that nanobubbles with width about 100 nm were obtained on the sample surface. To conduct coalescence experiment, we first apply 80% setpoint to scan the area. After that, the area was rescanned with 96% setpoint. The resulting image is shown in **Figure S4(b)**. One can see that nanobubbles with much larger size were obtained with reduced density. If the spherical objects were solid particles or blisters, they could not be coalesced in this way. Therefore, one can conclude that the spherical-cap like objects were neither solid particles, nor blisters.

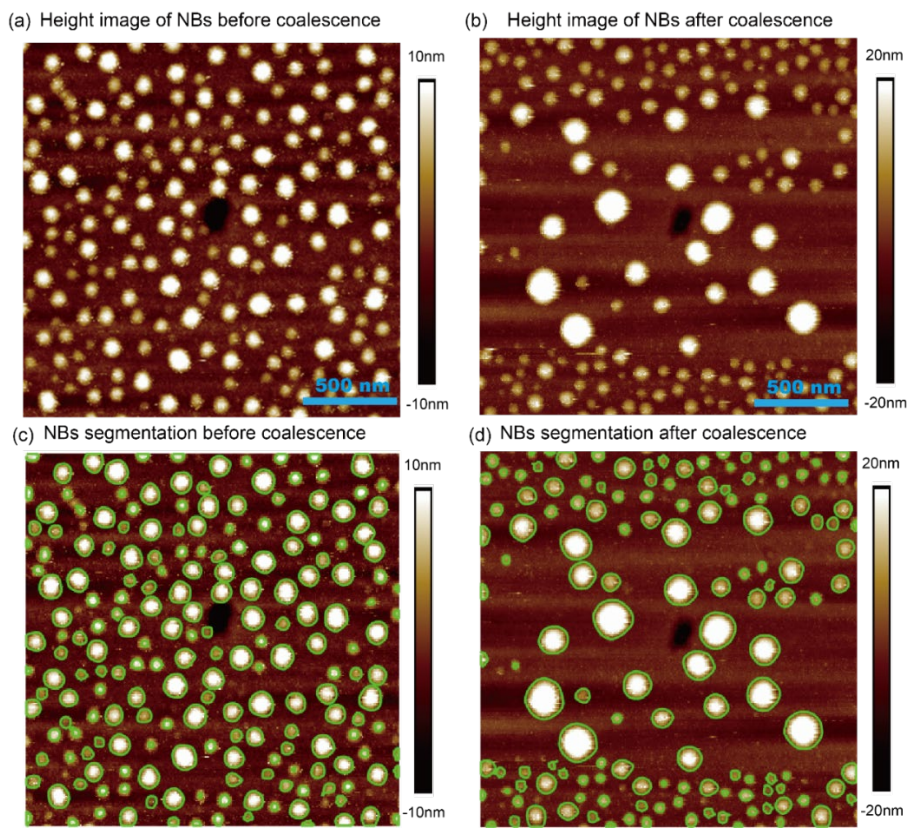

*Figure S4. Nanobubble coalescence performed on a PS surface. (a) AFM height image of the sample surface after immersed into DI water. (b) AFM height image of the same area after a higher scan load was applied. Nanobubbles at lower number density, but much larger size were obtained, indicating nanobubble coalescence. (c-d) NB image segmentation for NB height images in (a) and (b), respectively. The green contours are detected NB boundaries with a homemade NB segmentation algorithm.*

Moreover, the NB volumes before and after the coalescence were measured with a home-made NB image segmentation algorithm [15]. To do so, the AFM images of the NBs were first segmented. The results are shown in **Figure S4(c)** and **S4(d)** for NB images shown in **Figure S4(a)** and **S4(b)**, respectively. With the segmented NB images, the NB volumes were calculated. The total volumes of NBs in **Figure S4(a)** and **S4(b)** are about $3.7 \times 10^6$ nm$^3$ and $7.1 \times 10^6$ nm$^3$, respectively. The obtained result is in a good agreement with what we reported recently [15]. The increased NB volume after NB coalescence may partly be explained by the reduced Laplace pressure in the larger NBs, and partly be by diffusion of dissolved gas into the merging process. A quantitative analysis of these two effects is beyond the scope of the present paper.

*Tip-Sample Interaction Experiments*

The second approach to distinguish NBs from solid particles and blisters is an amplitude-distance curve measurement above the spherical-cap-like objects. During extension or retraction motion of an oscillating tip relative to a NB, the amplitude of the tip shows gradually decreases with reducing tip-sample separation distance [14, 16]. However, the oscillating amplitude of AFM tips rapidly decreases once they get into contact with surfaces of solid particles or blisters, very different from the behavior of nanobubbles. One example of the amplitude-distance curve of AFM obtained on a spherical-cap like object is given in **Figure S5**. **Figure S5(a)** is an AFM height image of the PS sample surface (sample 1). In the center area, there is a big spherical-cap-like object. We measured amplitude-distance curves on the object and obtained two kinds of curves. Here we took amplitude-distance curves obtained at two selected points (marked as 1 and 2). The results are shown in **Figure S5(b)**.

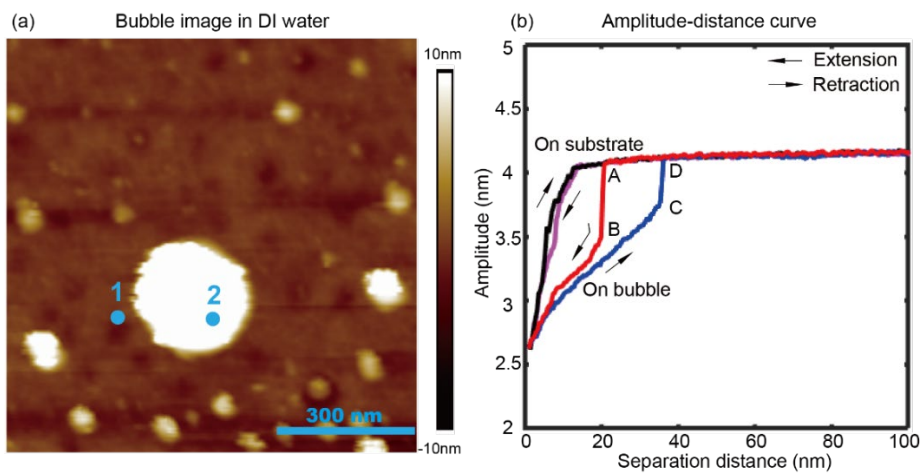

*Figure S5. Amplitude-distance curve measurement on a spherical-cap like object and sample substrate. (a) AFM height image of the sample surface in DI water, showing a big spherical-cap like object in the center. (b) Two typical amplitude-distance curves measured at point 1 and 2, respectively.*

One kind of curve is the one obtained above the substrate (point 1 in **Figure S5(a)**). Once the tip contacts the substrate, the amplitude rapidly decreases with decreasing tip-sample separation distance. The other kind of curve is the one obtained above the spherical-cap like domains (point 2 in **Figure S5(a)**). One can see that the obtained result is consistent with the one of our recent investigation [16]. As described there [16], there are typical snap-in period A → B and detached period C → D, which corresponds to the position where the AFM tip gets and loses contact with the object, respectively. The tip-sample interaction further confirms that the obtained spherical-cap like objects were not solid ones, neither blisters.

*Degassing Experiments*

To distinguish NBs from liquid contamination (PDMS droplets, for example), we applied degassing experiments[12, 17, 18]. For a PS sample, we first imaged the sample surface in air. The image is shown in **Figure S6(a)**. One can see that the surface is flat without any spherical-cap like objects. After that, the sample was immersed into DI water. We first obtained spherical-cap like objects on the sample surface, as shown in **Figure S6(b)**. Then, the AFM liquid cell with the sample immersed in water was transferred into a vacuum chamber to degas the water for over 4 hours. After that, the liquid cell was brought back to AFM and was re-scanned. The result is shown in **Figure s6(c)**. One can see that the spherical-cap like objects had all disappeared. This implies that the observed spherical-cap like objects were actually gas bubbles and not droplets of a poorly solvable liquid.

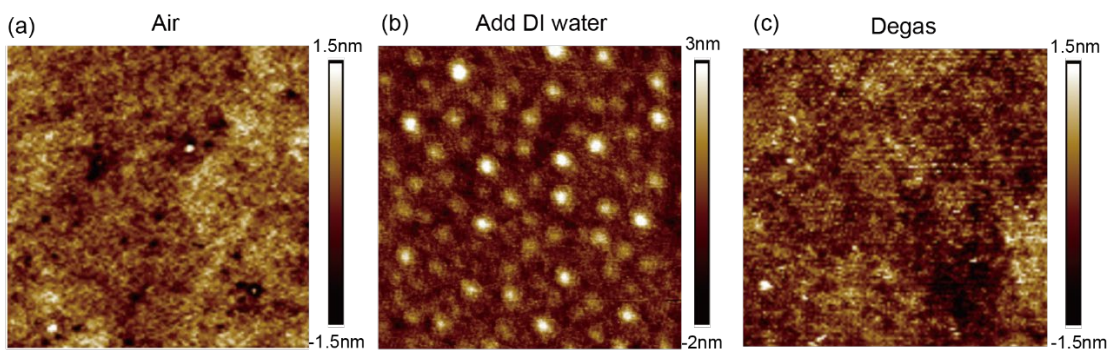

*Figure S6 Degassing experiment. (a) A PS sample surface obtained with tapping mode AFM in air. (b) Spherical-cap like objects obtained on the sample surface after it was immersed into DI water. (c) AFM height image of the same sample after the liquid cell (the sample remained in water during degassing) was kept for 4 hours in a vacuum chamber for degassing. The NBs have dissolved.*

## Deconvolution of Tip Radius and Correction of NB Contact Angles

It is well known that a topography image obtained from an AFM is actually the convolution of the AFM tip and substrate morphologies [19-21]. Therefore, the measured NB width needs to be corrected due to the finite size of the cantilever tip radius. In the case of the spherical-cap-liked NBs, the influence of the AFM tip on the contact angles of the protruding NBs is illustrated in **Figure S7a**. In the figure, the dashed cross section is the directly measured one from AFM images, while the solid one is the NB's actual cross section. From the figure, one can see that the tip radius causes overestimation of the NB width. The deconvolution of the AFM tip gives a corrected expression of the NB radius, width, and contact angle, namely[19-21]

$$R'_c = \frac{(D)^2 + 4H^2}{8H} - R_{tip}, \tag{S1}$$

$$D' = \sqrt{D^2 - 8HR_{tip}}, \tag{S2}$$

and

$$\theta' = 2\arctan(2H/D'), \tag{S3}$$

where $D'$ and $D$, $\theta'$ and $\theta$, $R_c'$ and $R_c$, $H'$ and $H$ ($H' = H$) are the actual and apparent (uncorrected) values of width, contact angles, radius of curvature, and height of the NB, respectively, and $R_{tip}$ is the tip radius (about 8 nm in this case).

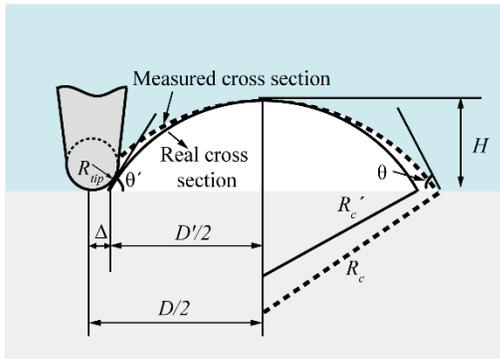
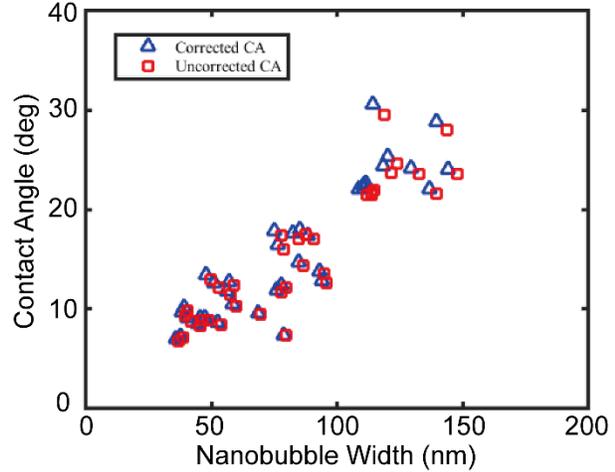

*Figure S7* *(a) Schematic sketch of the NB size correction through deconvolution of the tip radius. (b) Comparison of the corrected and uncorrected contact angle (CA) for protruding NBs. The difference between the corrected contact angle and uncorrected ones is slight.*

With **Eq. (S3)**, NB contact angle can be corrected. For the NBs with width of 50-150 nm, we found that $D$ is about 2-4 nm larger than $D'$. A comparison of the corrected and uncorrected contact angle of NBs is conducted to evaluate the effect of finite size of tip radius on contact angle measurement, as shown in **Figure S7b**. From the result, one can see that the uncorrected CA is slightly lower (about 2%) than that of the corrected one.

The classification of planar and protruding NBs is based on the value of their contact angles.

The distribution of contact angles for all trapped NBs is shown in **Figure S8**. In this study, the NBs with contact angles of 0º or even slightly less are referred to as planar NBs. The others are protruding NBs, of which contact angles are normally larger than 10 º.

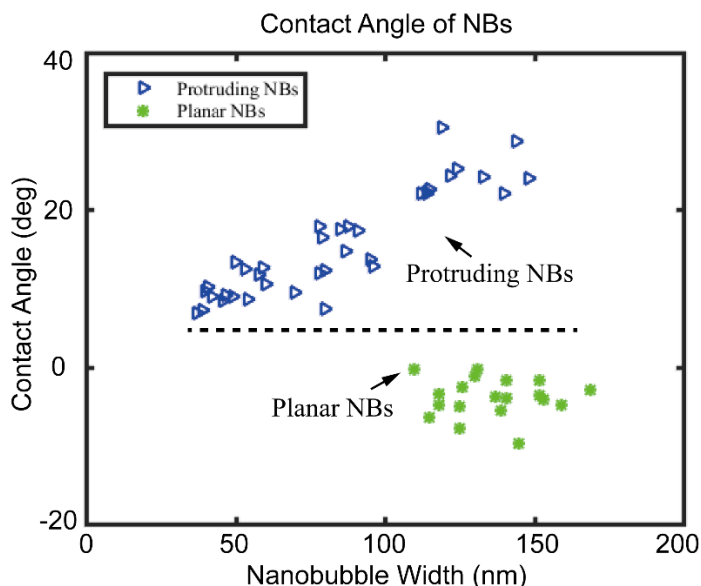

*Figure S8* Corrected contact angles for all trapped NBs (planar and protruding). The contact angle for planar NBs is mostly less than 0º. The dashed horizontal line in the figure divides the trapped NBs into two categories: protruding and planar NBs.

**References:**

1.   Y. Wang, B. Bhushan and X. Zhao, *Langmuir*, 2009, **25**, 9328-9336.
2.   G. Reiter, *Langmuir*, 1993, **9**, 1344-1351.
3.   Y. Wang, B. Bhushan and X. Zhao, *Nanotechnology*, 2009, **20**, 045301.
4.   H. Tarabkova and P. Janda, *Journal of Physics-Condensed Matter*, 2013, **25**, 184001.
5.   P. Janda, O. Frank, Z. Bastl, M. Klementova, H. Tarabkova and L. Kavan, *Nanotechnology*, 2010, **21**, 095707.
6.   J. H. Weijs, B. Andreotti and J. H. Snoeijer, *Soft Matter*, 2013, **9**, 8494-8503.
7.   H. R. Brown and T. P. Russell, *Macromolecules*, 1996, **29**, 798-800.
8.   H. Meyer, T. Kreer, A. Cavallo, J. P. Wittmer and J. Baschnagel, *Eur. Phys. J.-Spec. Top.*, 2007, **141**, 167-172.
9.   L. Si, M. V. Massa, K. Dalnoki-Veress, H. R. Brown and R. A. L. Jones, *Phys. Rev. Lett.*, 2005, **94**, 127801.
10.  J. H. Lee, J. Y. Chung and C. M. Stafford, *Acs Macro Letters*, 2012, **1**, 122-126.
11.  D. Lohse and X. Zhang, *Rev. Mod. Phys.*, 2015, **87**, 981-1035.
12.  R. P. Berkelaar, E. Dietrich, G. A. M. Kip, E. S. Kooij, H. J. W. Zandvliet and D. Lohse, *Soft Matter*, 2014, **10**, 4947-4955.
13.  R. P. Berkelaar, P. Bampoulis, E. Dietrich, H. P. Jansen, X. H. Zhang, E. S. Kooij, D. Lohse and H. J. W. Zandvliet, *Langmuir*, 2015, **31**, 1017-1025.
14.  B. Bhushan, Y. Wang and A. Maali, *J. Phys.: Condens. Matter*, 2008, **20**, 485004.
15.  Y. Wang, H. Wang, S. Bi and B. Guo, *Beilstein Journal of Nanotechnology*, 2015, **6**, 952-963.
16.  Y. Wang, H. Wang, S. Bi and B. Guo, *Scientific Reports*, 2016, **6**, 30021.

17. L. Bao, Z. Werbiu, D. Lohse and X. Zhang, *J. Phys. Chem. Lett.*, 2016, **7**, 1055-1059.
18. L. J. Zhang, Y. Zhang, X. H. Zhang, Z. X. Li, G. X. Shen, M. Ye, C. H. Fan, H. P. Fang and J. Hu, *Langmuir*, 2006, **22**, 8109-8113.
19. B. M. Borkent, S. de Beer, F. Mugele and D. Lohse, *Langmuir*, 2010, **26**, 260-268.
20. D. Li and X. Zhao, *Colloid Surf. A-Physicochem. Eng. Asp.*, 2014, **459**, 128-135.
21. B. Song, W. Walczyk and H. Schoenherr, *Langmuir*, 2011, **27**, 8223-8232.